\documentclass[aps,superscriptaddress,preprint,nofootinbib]{revtex4}

\makeatletter

\usepackage{natbib}
\usepackage{dcolumn}
\usepackage{graphicx}
\usepackage{amsmath}
\usepackage[hang,small,bf,centerlast]{caption} 
\usepackage{rotating}
\begin{document}
\title{Statistically validated networks in bipartite complex systems}

\author{Michele Tumminello}
\affiliation{Dipartimento di Fisica e Tecnologie Relative, Universit\`a di Palermo, viale delle Scienze Ed.18, I-90128 Palermo, Italia}

\author{Salvatore Miccich\`{e}}
\affiliation{Dipartimento di Fisica e Tecnologie Relative, Universit\`a di Palermo, viale delle Scienze Ed.18, I-90128 Palermo, Italia}

\author{Fabrizio Lillo}
\affiliation{Dipartimento di Fisica e Tecnologie Relative, Universit\`a di Palermo, viale delle Scienze Ed.18, I-90128 Palermo, Italia}
\affiliation{Santa Fe Institute, 1399 Hyde Park Road, Santa Fe NM 87501, USA}

\author{Jyrki Piilo}
\affiliation{Turku Centre for Quantum Physics, Department of Physics and Astronomy, University of Turku, FI-20014 Turun yliopisto, Finland}

\author{Rosario N. Mantegna}
\affiliation{Dipartimento di Fisica e Tecnologie Relative, Universit\`a di Palermo, viale delle Scienze Ed.18, I-90128 Palermo, Italia}

\maketitle 

{\bf 
Many complex systems present an intrinsic bipartite nature and are often described and modeled in terms of networks \cite{Watts1998,barabasi1999,newman2002,song2005,schweitzer2009}. Examples include movies and actors \cite{Watts1998,barabasi1999,song2005}, authors and scientific papers \cite{newman2001,barabasi2002,guimera2005,colizza2006}, email accounts and emails \cite{McCallum2007}, plants and animals that pollinate them \cite{bascompte2003, reed2009}.  
Bipartite networks  are often very heterogeneous in the number of relationships that the elements of one set establish with the elements of the other set. When one constructs a projected network with nodes from only one set, the system heterogeneity makes it very difficult to identify preferential links between the elements. Here we introduce an unsupervised method to statistically validate each link of the projected network against a null hypothesis taking into account the heterogeneity of the system. We apply our method to three different systems, namely the set of clusters of orthologous genes (COG) in completely sequenced genomes \cite{Tatusov1997,Tatusov2003}, a set of daily returns of 500 US financial stocks, and the set of world movies of the IMDb database \cite{IMDb}. 
In all these systems, both different in size and level of heterogeneity, we find that our method is able to detect network structures which are informative about the system and are not simply expression of its heterogeneity.  Specifically, our method (i) identifies the preferential relationships between the elements, (ii) naturally highlights the clustered structure of investigated systems, and (iii) allows to classify links according to the type of statistically validated relationships between the connected nodes. }

Bipartite networks are composed by two disjoint sets of nodes such that every link connects a node in the first set with a node of the second set. The bipartite network is often transformed by one-mode projecting, i.e. one creates a network of nodes belonging to one of the two sets and two nodes are connected when they have at least one common neighboring node of the other set. 
Two main and related problems arise in these networks. First, there is often a large heterogeneity in the node degree of the bipartite network and the information associated to this heterogeneity is partially lost in the projected network. Second, a link in a projected network could indicate a preferential relationships between two specific nodes or could be consistent with the degree of heterogeneity of the system. In this letter we introduce a statistical method to validate the presence of a link by simultaneously coping with the heterogeneity of the system.
 
A bipartite system presents two sources of heterogeneity associated with the two sets. To be specific consider the example of the COG database \cite{Tatusov1997,Tatusov2003}, where one set is composed by 66 organism's genomes and the other by 4,873 COGs. The number of COGs in a genome is heterogeneous, ranging from 362 to 2,243. Similarly, the number of genomes in which a specific COG is present ranges from 3 to 66. Here we consider the projected network of genomes. A COG$_k$ is a COG present in $k=3,\cdots,66$ genomes. 
In each subset of COG$_k$s we are therefore left only with the heterogeneity of organisms. We identify a preferential relationship between each pair of organisms by statistically validating the co-occurrence of COG$_k$s against a null hypothesis that takes into account such heterogeneity. 
Specifically, given two organisms  A and  B,  let $N_A$ be the number of COG$_k$s in organism A and $N_B$ the number of COG$_k$s in organism B. The total number of COG$_k$s is $N_k$ and the observed number of COG$_k$s belonging to both A and B is $N_{AB}$. Under the null hypothesis of random co-occurrence, the probability of observing $X$ co-occurrences is given by the hypergeometric distribution \cite{Feller}
\begin{equation}
H(X|N_k,N_A,N_B)=\frac{{N_A \choose X} {N_k-N_A \choose N_B-X}}{{N_k \choose N_B}}.
\label{hyper}
\end{equation}
We can therefore associate a $p$-value to the observed $N_{AB}$ as $p(N_{AB}) =1- \sum_{X=0}^{N_{AB}-1} H(X|N_k,N_A,N_B)$.
It is worth noting that the described null hypothesis directly takes into account the heterogeneity of organisms with respect to the number of COGs present in their genome. 
For each pair of organisms, we separately evaluate the $p$-value for each subset of COG$_k$ and we count the number of subsets in which the $p$-value is smaller than a selected statistical threshold.  
We accordingly set a link between A and B if the number of subsets is non vanishing and we use it as the weight of the link. 
The described link validation procedure involves multiple hypothesis testing, across organism pairs and subsets of COG$_k$. Therefore the statistical threshold must be corrected for multiple comparisons. The most stringent method to address this problem is the Bonferroni correction \cite{Miller1981}. It is based on the consideration that if one tests $N_t$ either dependent or independent hypotheses on a set of data, then a conservative way of maintaining the error rate low is to test each individual hypothesis at a statistical significance level of $p_t/N_t$, where $p_t$ is the chosen statistical threshold ($1 \%$ in the present study). In our case the number of organisms is $N_o=66$ and we test $N_t=64 N_o (N_o-1)/2$ hypotheses, equal to the number of pairs of organisms times the number of subsets of COG$_k$. Thus our Bonferroni threshold is $p_b=0.01 \cdot 2 /(64 N_o (N_o-1))\cong 7.3\times 10^{-8}$.  We refer to the network obtained by using the Bonferroni threshold as the Bonferroni network. 
A less stringent correction for multiple comparisons is the False Discovery Rate (FDR) \cite{Benjamini1995}. The threshold of the FDR correction linearly increases with the number of tests in which the null hypothesis is rejected. We refer to the network obtained by using the FDR correction for multiple comparisons as the FDR network. By construction, the Bonferroni network is a subgraph of the FDR network, which is a subgraph of the adjacency network.  

\begin{figure}
\includegraphics[scale=0.3,angle=0]{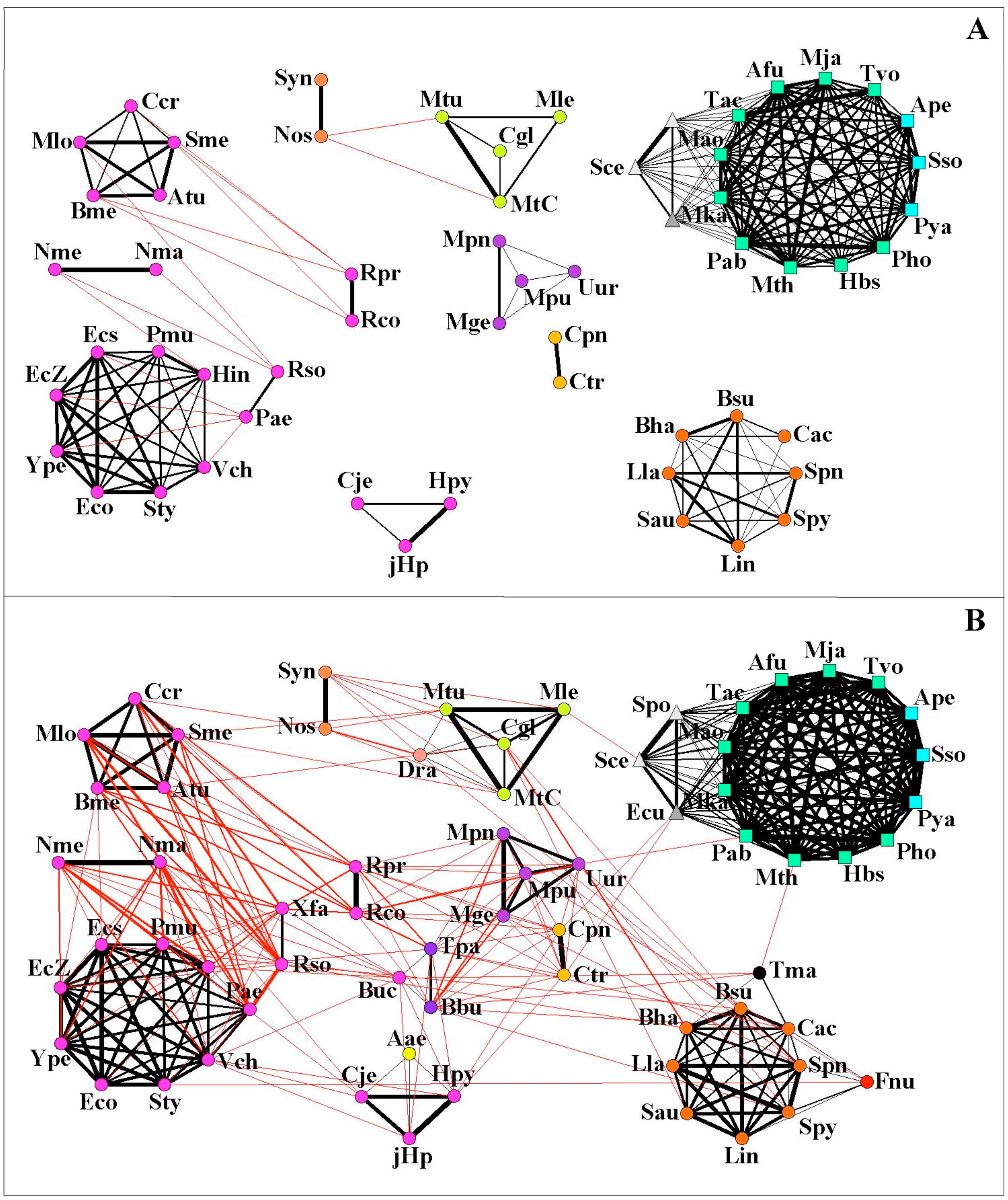}
\caption{Bonferroni (Panel A) and FDR (Panel B) networks of the organisms investigated in the COG database. The shape of the node indicates the super kingdom of the organism: Archaea (squares), Bacteria (circles), and Eukaryota (triangles). The color of the node indicates the phylum of the organism. The thickness of the link is related to its weight and is proportional to the logarithm of the number of COG$_k$ validations between the two connected nodes. Red links are those removed when Infomap \cite{Rosvall2008} is applied and thus they connect different communities of organisms. 
The Bonferroni network (Panel A) presents 7 connected components plus 8 isolated nodes (not shown). The largest connected component (on the left) is composed of bacteria belonging to the phylum of Proteobacteria. Subgroups belonging to different classes can also be recognized. In fact, Eco, Ecz, Ecs, Ype, Hin, Pmu, Vch, Pae and Sty belong the class of Gammaproteobacteria whereas Atu, Sme, Bme, Ccr, Rpr, Rco and Mlo  are Alphaproteobacteria and NmA, Nme and Rso are Betaproteobacteria. 
The second connected component is composed by Archaea genomes belonging to the two phyla of Euryarchaeota (Mth, Mja, Hbs, Tac, Tvo, Pho, Pab, Afu, Mka, and Mac) and Crenarchaeota (Pya, Sso and Ape). Interestingly Archaea are also linked to the three unicellular eukaryotes present in the set (Ecu, Sce and Spo) although the weight of the links of eukariotes with Archaea is markedly smaller than the one observed for links occurring among Archaea \cite{ciccarelli2006}.  The FDR network (Panel B) is connected. The group of Archaea and Eukaryota is clearly distinct from the network region of Bacteria.}
\label{FDRBONF} 
\end{figure}

The Bonferroni network of organisms includes 58 non isolated nodes connected by 216 weighted links (Fig.~\ref{FDRBONF}A) and it shows seven connected components, each one having a clear biological interpretation in terms of organisms' lineage. The FDR network of organisms includes all the 66 organisms and the number of weighted links in this network is 369 (Fig.~\ref{FDRBONF}B). Thus the entire set is covered and the additional links provides relations among the groups already observed in the Bonferroni network. Note that the adjacency network of this system is a complete graph. 

It is worth noting that both the Bonferroni and the FDR network display a clear cluster structure and the clusters have a direct biological interpretation in terms of lineages. Even if the FDR network is completely connected, the application of community detection algorithms \cite{Fortunato2010}, such as Infomap \cite{Rosvall2008}, to the statistically validated networks gives a clear community structure (see Fig.~\ref{FDRBONF}). This is not true for the adjacency network and shows that the statistically validated networks are able to identify the many relevant links inside communities and the few relevant links between different communities of organisms.

As a second example we consider the collective dynamics of the daily return of $N_s=500$ highly capitalized US financial stocks in the period 2001-2003 (748 trading days). The two sets of the bipartite system are the stocks and the days, and we consider here the projected network of stocks. The interest in this example is that we (i) generalize our procedure to complex systems where the elements are monitored by continuous variables, (ii) show how to simplify the above procedure when the second source of heterogeneity (in the above example the COG frequency in different organisms) is small, and (iii) show how to classify links according to the type of relation between the two nodes.

Since we want to identify similarities and differences among stock returns not due to the global market behavior, we investigate the excess return of each stock $i$ with respect to the average daily return of all the stocks in our set.
The excess return of each stock $i$ at day $t$ is then converted into a categorical variable with 3 states: \emph{up}, \emph{down}, and \emph{null}. For each stock we introduce a daily varying threshold $\sigma_i(t)$ as the average of the absolute excess return (a measure of local volatility) of stock $i$ over the previous 20 days. State \emph{up} (\emph{down}) is assigned when the excess return of stock $i$ at day $t$ is larger (smaller) than $\sigma_i(t)$ (-$\sigma_i(t)$). The state \emph{null} is assigned to the remaining days. 
We study the co-occurrence of states \emph{up} and \emph{down} for each pair of stocks. 
In this case we can neglect the heterogeneity of state occurrence in different trading days because the number of \emph{up} (\emph{down}) states is only moderately fluctuating across different days and it has a bell shaped distribution with a range of fluctuations smaller than one decade for each stock. With this approximation we can statistically validate the co-occurrence of state $A$ (either \emph{up} or \emph{down}) of stock $i$ and state $B$ (either \emph{up} or \emph{down}) of stock $j$ with the following procedure. Let us call $N_A$ ($N_B$) the number of days in which stock $i$ ($j$) is in the state $A$ ($B$). Let us call $N_{AB}$ the number of days when we observe the co-occurrence of state $A$ for stock $i$ and state $B$ for stock $j$. Under the null hypothesis of random co-occurrence of state $A$ for stock $i$ and state $B$ for stock $j$, the probability of observing $X$ co-occurrences of the investigated states of the two stocks in $T$ observations is again described by the hypergeometric distribution, $H(X|T,N_A,N_B)$. As before we can associate  $p$-value with each pair of investigated stocks for each combination of the investigated states. We indicate the state \emph{up} (\emph{down}) of stock $i$ as $i_{u}$ ($i_{d}$). The possible combinations are ($i_{u}$,$j_{u}$), ($i_{u}$,$j_{d}$), ($i_{d}$,$j_{u}$), and ($i_{d}$,$j_{d}$).
As before the statistical test is a multiple hypothesis test and therefore either the Bonferroni or FDR correction is necessary. The Bonferroni threshold is $p_b=p_t/(2 N_s (N_s-1))$ where the denominator of the threshold is the number of considered stock pairs ($N_s (N_s-1)/2$) times 4, which is the number of different co-occurrences investigated. 
Each pair of stocks is characterized by the set of the above four combinations which are statistically validated. There are $2^4-1=15$ possible cases (relationships) with at least one validation, but we observe only 5 cases: L1 in which the co-occurrences ($i_{u}$,$j_{u}$) and  ($i_{d}$,$j_{d}$) are both validated; L2 in which only the co-occurrence ($i_{d}$,$j_{d}$) is validated, L3 in which only the co-occurrence ($i_{u}$,$j_{u}$) is validated, L4 in which either only ($i_{u}$,$j_{d}$) or only ($i_{d}$,$j_{u}$) is validated;  and L5 when both the co-occurrence ($i_{u}$,$j_{d}$) and ($i_{d}$,$j_{u}$) are validated. Note that we put in the same relationship L4 two cases which are different only for the order in which the two nodes are considered.
The set of relationships $L1$, $L2$, and $L3$ and the associated links describe a coherent movement of the price of the two stocks, while the set of relationships $L4$ and $L5$ describes opposite deviation from the average market behavior. We can therefore construct networks where the statistically validated links are associated with a label that specifies the type of relationship between the two connected nodes. This structure is richer than a simple unweighted network, but it is also different from a weighted network because it describes relationships which cannot be described by a numerical value only. We address the set of different relationships present between two nodes of the statistically validated network with the term multi-link. 

\begin{figure}
\includegraphics[scale=0.3,angle=0]{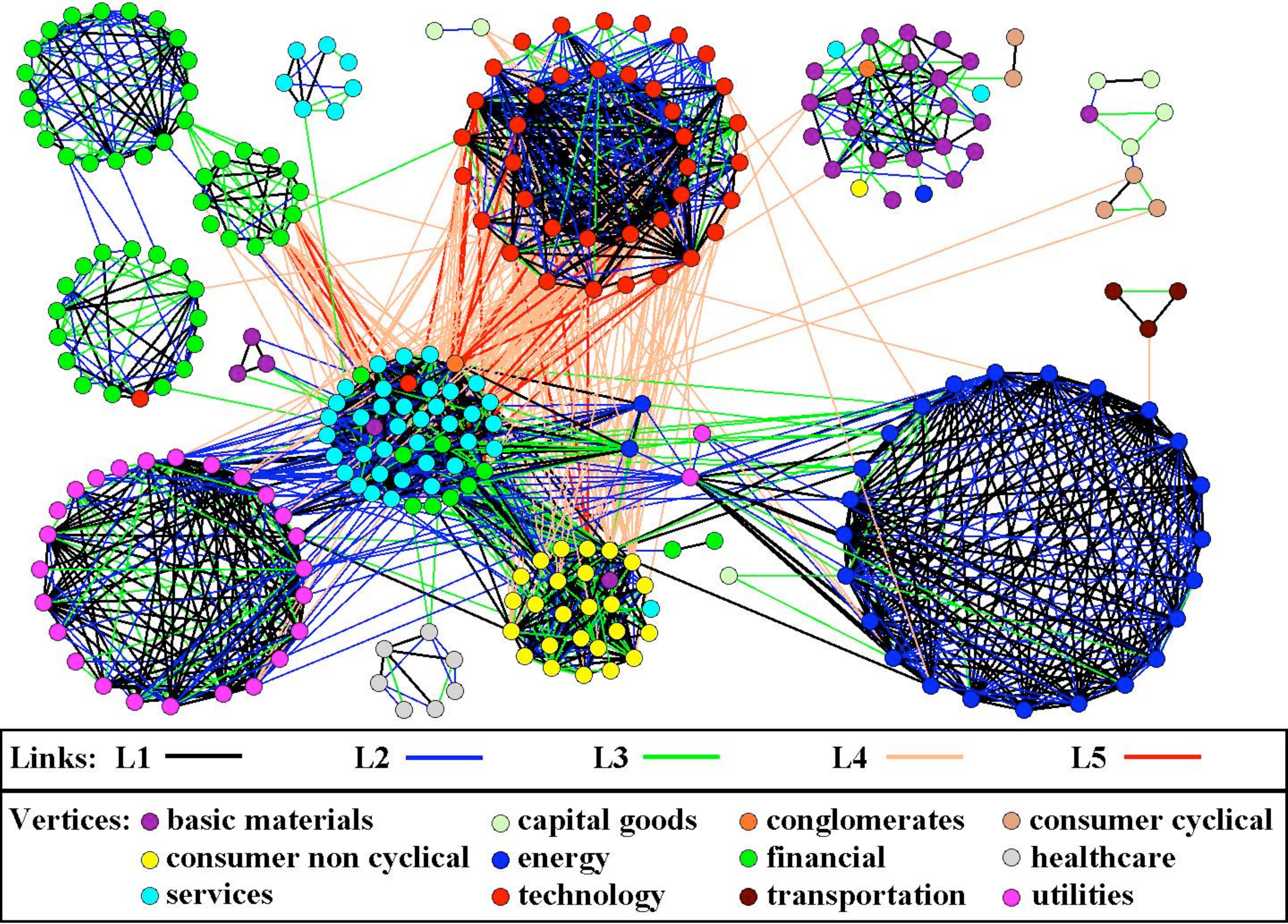}
\caption{The largest connected component of the Bonferroni network associated with the system of 500 stocks. The nodes represent stocks and links connecting different stocks correspond to the statistically validated relationships. The node color identifies the economic sector of the corresponding stock. The economic sector classification is done according to Yahoo Finance. The color of a multi-link identifies the corresponding validated relationship.}
\label{bonferroni500cc} 
\end{figure}

The Bonferroni network of the system is composed by 349 stocks connected by 2,230 multi-links.  The multi-links are of different nature. Specifically, we observe 1,158 $L1$-links, 494 $L2$-links, 354 $L3$-links, 196 $L4$-links, and 28 $L5$-links. The largest connected component of the network includes 273 stocks. There also are 19 smaller connected components of size ranging from 2 to 15.
In Fig.~\ref{bonferroni500cc} we show the largest connected component of the Bonferroni network.  
It presents several regions in which stocks are strongly connected by $L1$, $L2$, and $L3$ multi-links. These regions are very homogeneous in the economic sector of the stocks. The connection between different regions is mostly provided by  a large number of $L4$ and $L5$ multi-links. This is especially evident for the group of technology stocks (red circles) for which all except one of the multi-links outgoing from the group are $L4$ or $L5$ multi-links, indicating moderate or strong anti-correlation of technology stocks with the other groups. The strongest anti-correlation is detected between technology and services stocks (cyan circles). A partition analysis of the Bonferroni network shows (see Supplementary Information) that the services stocks, which are  anti-correlated with the technology stocks, mostly belong to the economic sub-sector of Real estate operations. We have also computed the FDR network of the system. As expected, it includes more stocks (494) and more multi-links (11,281), since the requirement on the statistical validation is less restrictive. The FDR network has a single connected component and the fraction of $L4$ and $L5$ multi-links is higher (35.9 \%) than in the case of the Bonferroni network (10.0 \%).

As before the adjacency network of stocks is a complete graph. On the contrary both the Bonferroni and the FDR networks display a highly clustered structure with clusters having a clear economic meaning. The use of Infomap on these statistically validated networks gives a partition in communities, which are extremely homogeneous by economic sector (see Supplementary Information). Therefore our method allows to construct networks where (i) links are statistically validated against a null hypothesis, (ii) multi-links describe qualitatively different relationships between pairs of stocks, e.g. both co-movements and opposite movements occurring between pairs of stocks, and (iii) a very accurate identification of communities of stocks is possible. To the best of our knowledge the presence of all these features is pretty unique and it is not shared by other similarity networks 
\cite{Tumminello2010} based on topological constraints \cite{Mantegna1999, Bonanno03, Tumminello2005},  correlation threshold \cite{Onnela2003}, or validated with bootstrap \cite{Tumminello2007}.

The last system we investigate is the bipartite system of movies and actors. We collect data from the Internet Movie Database (IMDb), which is the largest web repository of world movies. We consider here the projected network of  the 89,605 movies produced in the period 1990-2008. The set includes movies realized in 169 countries, and at least one genre is specified for each movie. The number of involved actors is 412,143. We choose this example because (i) it is a large system, (ii) it has a large heterogeneity both in movies and in actors, and (iii) it allows a sophisticated cluster characterization analysis based on the characteristics of the movie, namely  genre, language, country, etc..    

The actors heterogeneity ranges between 1 and 247 and it is so pronounced that there is no practical way to eliminate it when constructing statistically validated networks of movies. The approach of the $k$-subsets is not feasible due to lack of sufficient statistics. Therefore, an approximate statistical validation can be performed, by taking into account only the movies heterogeneity. In the presence of this limitation a number of false positive links can be expected. In spite of this limitation, the results obtained for the statistically validated networks are very informative about several aspects of the movie industry. 

We construct the statistically validated networks of movies by testing the co-occurrence of actors in the cast of each movie pair. The null hypothesis we use is again given by the hypergeometric distribution, which naturally takes into account the heterogeneity due to the number of actors performing in each movie. Table \ref{summary} shows the severe filtering of nodes and links that is obtained in the validated networks of movies with respect to the adjacency network. Only 16\% (47\%) of the nodes and 1\% (7\%) of the links of the adjacency network are statistically validated in the Bonferroni (FDR) network. 

\begin{figure}
\includegraphics[scale=0.35,angle=0]{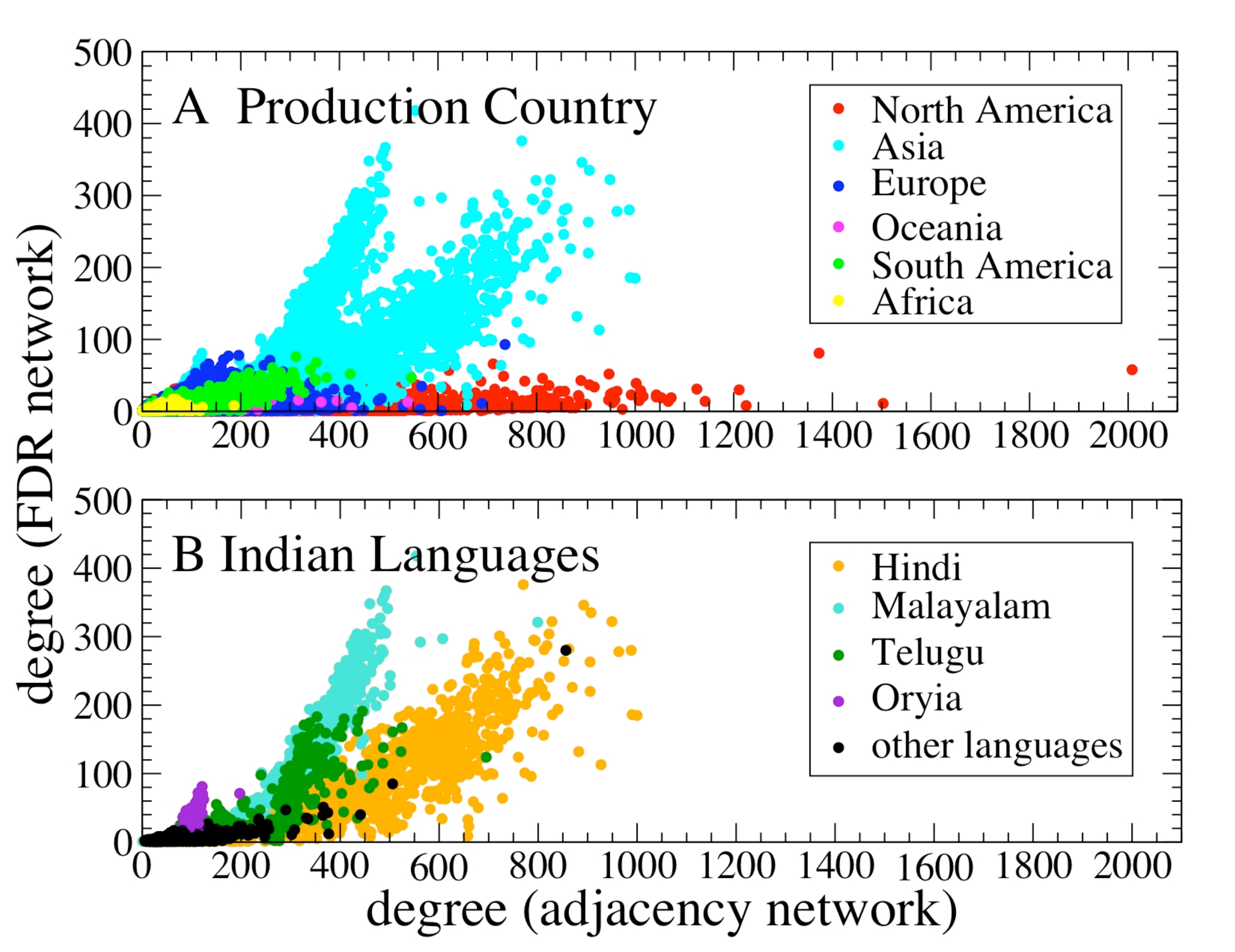}
\caption{Scatter plots of the degree of movies in the adjacency and FDR networks. Each circle represents a movie. We do not report movies with vanishing degree in at least one of the two networks. The panel A shows movies produced all over the world. The color of each symbol identifies the continent of the production country. Only movies with a single production country are shown. The panel B shows the data for the Indian movies and the color indicates the movie language. Only movies with a single language are shown. The North American movie industry  shows  typically a high degree of movies in the adjacency network and a relatively low degree in the FDR network, indicating a tendency to avoid a similar cast in different movies. A different behaviour is observed in Asia, while Europe is an intermediate case.  The Indian data shows that the behaviour is different within regions using different languages. A regional clustering of the Indian movie industry is efficiently detected by the statistically validated networks (see Supplementary Information).}
\label{fig:degdeg} 
\end{figure}

A comparison of the degree of movies in the adjacency and FDR networks allows to clearly  distinguish Asian movie industry from the rest of the world movie industry, and languages within single countries like India (see Fig.\ref{fig:degdeg}). According to the present state of the IMDb database, this comparison suggests that the Asian movie industry, and the Indian movie industry in particular, present a level of variety in their cast formation that is lower than the variety observed in the western movie industry conditioned to the number of actors present in the movie casts.   

In the Supplementary Information we analyze the movie communities detected when Infomap is applied to different networks. In the community detection of adjacency and statistically validated networks we
also weight links according to Ref.~\cite{newman2001b} to heuristically take into account actors' heterogeneity in performing movies. We find that, both in weighted and unweighted networks, the clusters of movies obtained from the Bonferroni and FDR networks have a high homogeneity in terms of production country, language, genre, and filming location. 

In summary, our  method allows to validate links describing preferential relationships among the heterogeneous elements of complex systems with intrinsic bipartite nature. Our method is very robust with respect to the presence of links, which are false positive. In fact, for all the investigated systems we verified that by randomly rewiring the bipartite network the associated Bonferroni network was empty. By applying the method to three different systems, we showed that it is extremely flexible since it can be applied to systems with different degree of heterogeneity and to systems described by binary, categorical, or continuous variables. 

\begin{table}
\caption{Basic properties of movie networks, namely the number of nodes, the number of links, the number of connected components, and the size of the largest connected component for the adjacency network and for the statistically validated networks.  The size of the smallest connected component is 2 for all networks. 
The largest connected component is covering almost completely the adjacency network, the largest fraction of movies in the FDR network (83 \%), but only 13 \% of the movies of the Bonferroni network. This shows that the Bonferroni network is already providing a natural partitioning of the movies included in it, even without the use of a partition algorithm on it. }
\begin{tabular}{|l||c|c|c|c|}
\hline
~ & ~Movies~ & Links & Numb. comp. & Largest c.c.\\
 \hline
Adjacency& $78,686$ & $2,902,060$&$647$&$77,193$\\
FDR& $37,429$ & $205,553$&$2,443$&$30,934$\\
Bonferroni& $12,850$ & $29,281$&$2,456$&$1,627$\\
\hline
\end{tabular}
\label{summary}
\end{table}

{\bf Supplementary Information} is linked to this manuscript. 

{\bf Software} A R code computing statistically validated networks of bipartite systems is available from the authors on request.  

{\bf Acknowledgements} We thank S. Fortunato and J. Kert\'esz for fruitful discussions. J.P. acknowledge financial support by The Magnus Ehrnrooth Foundation and  the Vilho, Yrj\"o, and Kalle V\"ais\"al\"a Foundation.

\newpage
\begin{center}
{\bf{SUPPLEMENTARY INFORMATION\\Statistically validated networks in bipartite complex systems}}
\end{center}

\section{Cluster detection and characterization}
In the present study we perform community detection\footnote{Girvan, M., \& Newman M.E.J., Community structure in social and biological networks {\em Proc. Natl. Acad. Sci. USA} {\bf 99}, 7821-7826 (2002).}$^,$\footnote{Fortunato, S., Community detection in graphs. {\em Physics Reports} {\bf 486}, 75-174 (2010).} on the adjacency and statistically validated networks, in order to put in evidence the different community structure of these networks. 

We obtain a partition of the vertices of networks by using the Infomap method by Rosvall and Bergstrom\footnote{Rosvall, M. \&  Bergstrom, C. T., Maps of random walks on complex networks reveal community structure. {\em Proc. Natl. Acad. Sci. USA} {\bf 105}, 1118-1123 (2008)}. This algorithm is considered one of the best$^2$ available today and it allows to efficiently investigate both weighted and unweighted networks. The method uses the probability flow of random walks in the considered network to identify the community structure of the system. This approach implies that two independent applications of the method to the same network may produce (typically slightly) different partitions of vertices. For each investigated network, we run the Infomap $10^3$ times and we select the best partition according to the minimal ``code length"$^3$. The obtained partition depends on whether the network is weighted or not, and eventually on how weights are selected. Therefore, in the following we discuss case by case the way in which cluster detection is performed. Once clusters of elements are detected, there still remains the problem of cluster interpretation. We address this problem by comparing the partition of the system as produced by the Infomap with an \emph{a priori} classification of the elements of the system. For instance movies can be characterized by their genre, and stocks can be characterized according to their economic sector. 

\subsection{Cluster characterization} 
Let us consider a system of $N$ elements and a specific detected cluster $C$ of $N_C$ elements to be characterized. Each element of the system has a certain number of attributes according to the considered \emph{a priori} classification, e.g. a movie can be classified as ``thriller" and ``drama". We indicate the total number of different attributes over all the elements of the system with $N_A$. For each attribute $Q$ of the system, e.g. the Financial sector for stocks or the genre Comedy for movies, we test if $Q$ is over-expressed in cluster $C$. In other words, we test if the number $N_{C,Q}$ of elements in cluster $C$ that have the attribute $Q$ is larger than what expected by randomly selecting the $N_C$ elements in the cluster from the total $N$ elements of the system. The probability that $X$ elements in cluster $C$ have the attribute $Q$, under the null hypothesis that elements in the cluster are randomly selected, is given by the hypergeometric distribution $H(X|N,N_C,N_Q)$, where $N_Q$ is the total number of elements in the system with attribute $Q$. Therefore, we can associate a $p$-value with the observed number $N_{C,Q}$ of elements in cluster $C$ that are classified with the attribute $Q$ according to the equation
\begin{equation}
\label{cluschar}
p(N_{C,Q}) =1- \sum_{X=0}^{N_{C,Q}-1} H(X|N,N_C,N_Q).
\end{equation}
If $p(N_{C,Q})$ is smaller than a given statistical threshold $p_b$ we say that the attribute $Q$ is over-expressed in cluster $C$, and therefore the attribute $Q$ is a characterizing aspect of cluster $C$. We separately test all the possible $N_A$ attributes for each detected cluster $C$. So, also in this case, we perform multiple comparisons, and we use a statistical threshold $p_b$ corrected for multiple comparisons by using the Bonferroni correction, i.e. we set $p_b=0.01/N_A$.

A pretty similar approach to the one described in this subsection is used in Gene Ontology analysis of gene expression profile\footnote{Dr$\breve{a}$ghici, S. {\it Data Analysis Tools for DNA Microarrays.} (Chapman and Hall/CRC, Boca Raton, 2003)}. 
\section{500 stocks: Community detection in multi-link statistically validated networks}

\subsection{Cluster detection}

Our multi-link statistically validated network of 500 stocks is a new kind of network presenting different kinds of links. For this reason it is \emph{a priori} not obvious how to proceed to detect communities on this network. Here we propose a minimalist approach distinguishing between co-occurrences of correlated evolution from co-occurrences of anti-correlated evolutions. Our procedure is as follows: in the community search on our network we remove all the links describing anti-correlated evolutions ($L4$ and $L5$), or, equivalently, we weight them with a zero weight. On the other hand, we weight the remaining links by taking into account whether the statistical validation of the link is single or twofold. With this choice,  the twofold links $L1$ have the weight equal to 2, whereas onefold links $L2$ and $L3$ have the weight equal to 1. While our approach is pragmatic and heuristic, we are aware that  a more theoretically based approach partitioning multi-link networks would certainly be useful in our approach and in the study of many other networks, where links of different nature can be naturally defined.

\subsection{Cluster characterization}

We analyze the clusters of stocks detected in the statistically validated networks by considering the over-expression of specific economic sectors and subsectors of stocks in each cluster. Each stock of the system is characterized by its economic sector\footnote{Economic sectors according to Yahoo Finance classification of stocks: Basic Materials, Capital Good, Conglomerates, Consumer Cyclical, Consumer Non Cyclical, Energy, Financial, Healthcare, Services, Technology, Transportation, Utilities.}, e.g. technology, financial, energy, etc. The total number of economic sectors is 12. Economic subsectors represent a more detailed classification of stocks. However, information about the economic subsector is missing for some stocks. For these stocks we simply keep the information about their economic sector also at this more detailed level of classification. There are $81$ different subsectors characterizing the $N=349$ non isolated stocks in the Bonferroni network, while there are $96$ subsectors  characterizing the $N=494$ non isolated stocks in the FDR network.
The characterization of clusters is separately done for the classification of stocks according to their sector ($p_b=0.01/12= 8\,10^{-4}$ in both the validated networks) and for their classification in terms of subsectors ($p_b=0.01/81= 1.2\,10^{-4}$ for the validation of clusters in the Bonferroni network and $p_b=0.01/96= 1.0\,10^{-4}$ for the validation in the FDR network). 

\begin{figure}
\includegraphics[scale=0.32,angle=0]{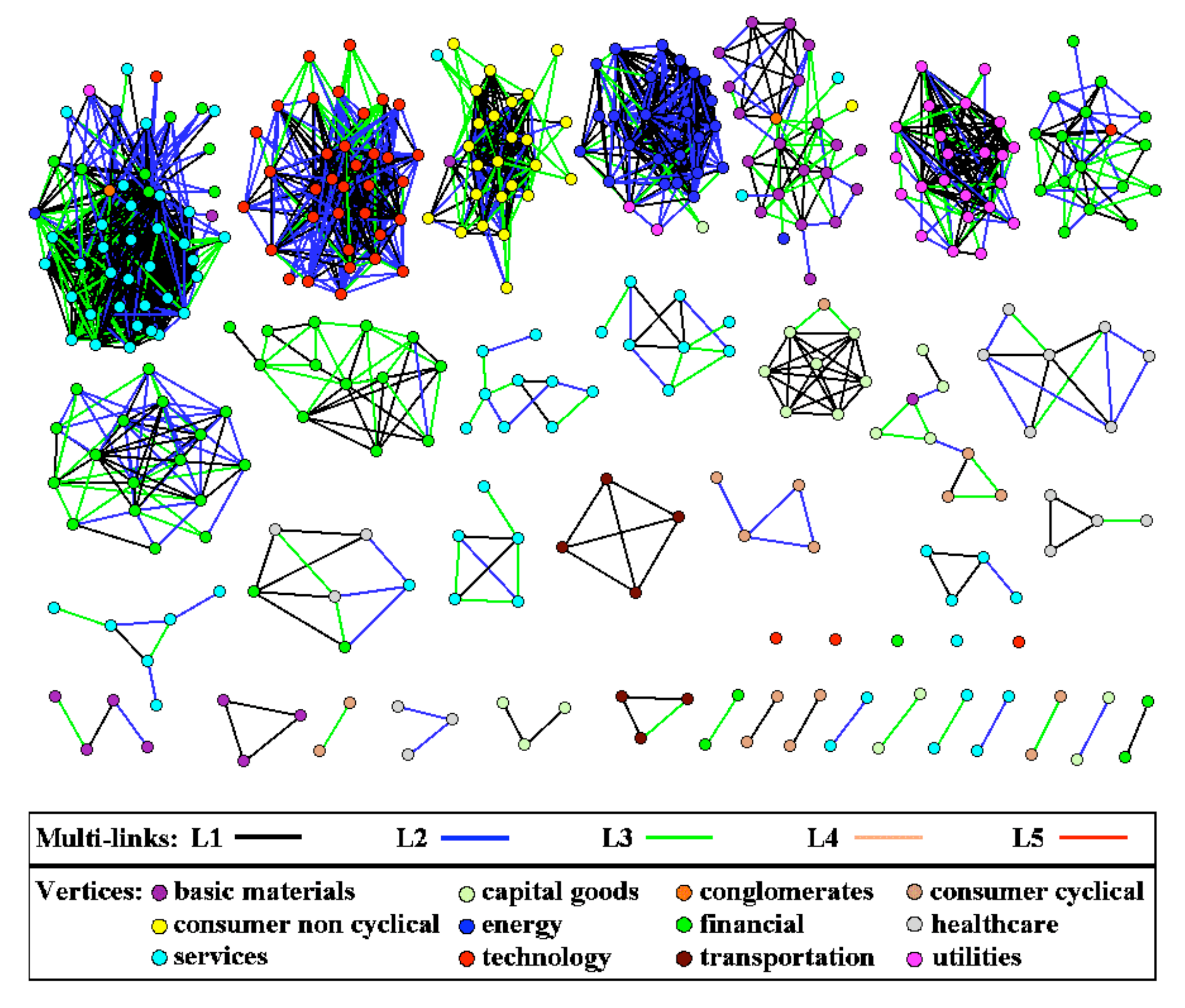}
\caption{Clusters of the Bonferroni network of 500 stocks traded in the US equity markets.  The seven clusters on the top row from left to right can be labelled by the over-expressions of economic subsectors as follows: 1) Services -- Real estate operations, 2) Technology -- Communication equipment, Technology -- Computer hardware, Technology -- Electronic instruments and control, and Technology -- Semiconductors, 3) Consumer -- Non-cyclical food processing, and Consumer --  Non-cyclical personal and household products, 4) Energy -- Oil and gas integrated, Energy -- Oil and gas operations, and Energy -- Oil well services and equipment, 5) Basic materials -- Chemical manufacturing, and Basic materials -- Chemical plastic and rubber, 6) Utilities --Electric utilities, and Utilities -- Natural gas utilities, and 7) Financial -- Insurance life, and Financial -- Insurance property and casualty.
}
\label{bonferroni500clusters} 
\end{figure}

The Infomap method detects 37 clusters with size ranging from 2 to 48 in the Bonferroni network. These clusters are shown in Fig.~\ref{bonferroni500clusters}.
When we perform the characterization at the level of subsectors, we detect 41 over-expressions of 31 distinct clusters. The number of over-expressed subsectors per cluster is therefore 1.32. Most of the clusters are described by a single economic subsector. 

\begin{figure}
\includegraphics[scale=0.32,angle=0]{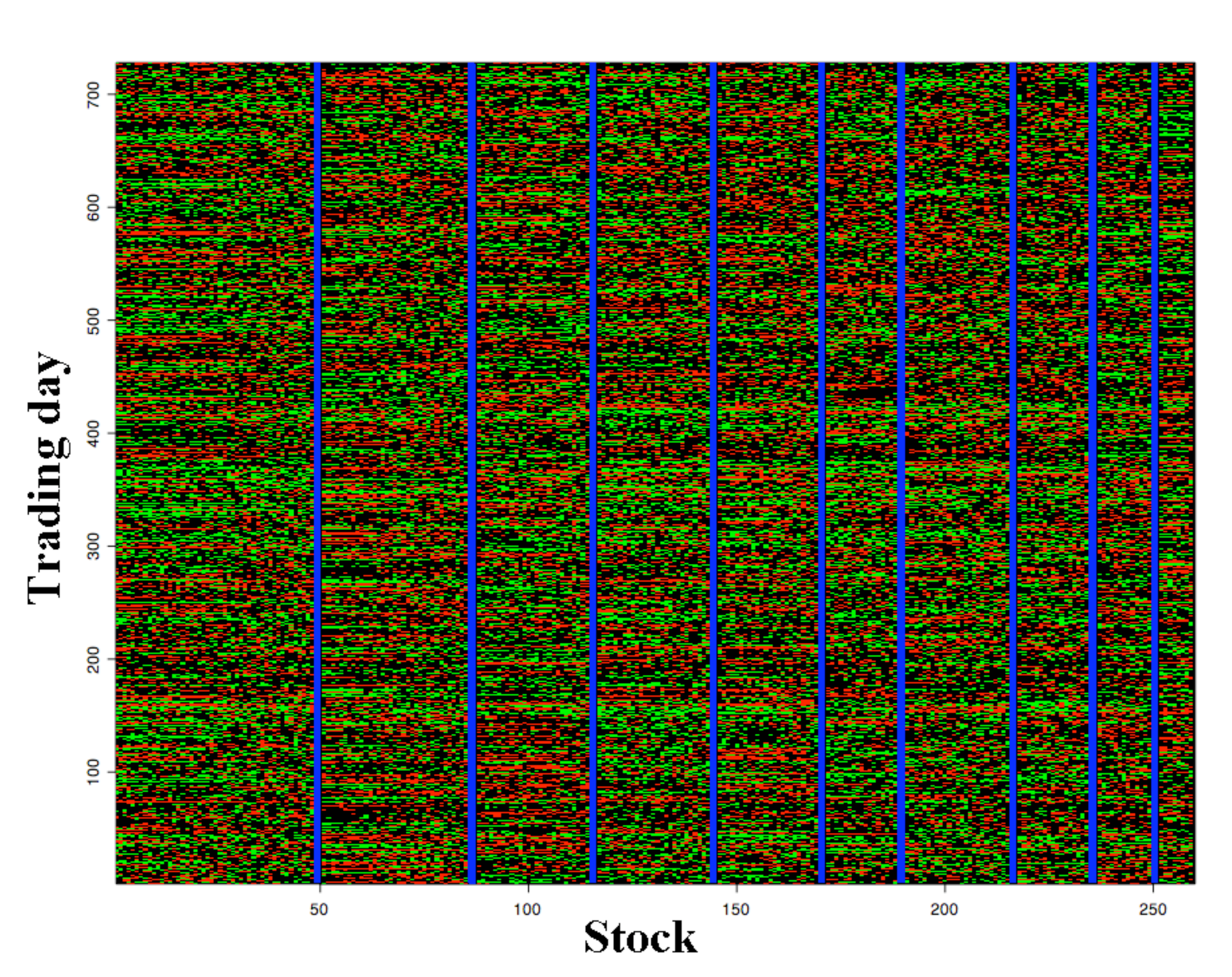}
\caption{Representation of the categorized excess return of stocks over time. The stocks belong to the 10 largest clusters of the Bonferroni network. State \emph{up} of the excess return is represented by a green spot, a red spot indicates the state \emph{down}, and a black spot the state \emph{null}. Stocks are ordered according to the partitioning in clusters and the order within a cluster is given according to the relevance of the stock in the cluster as detected by Infomap. Clusters of stocks are separated by a blue line. The clusters are from left to right: 1) Services -- Real estate operations, 2) Technology, 3) Energy, 4) Consumer -- non-cyclical, 5) Utilities, 6) Financial -- Regional banks, 7) Basic materials, 8) Financial -- Insurance, 9) Financial -- Investment services and 10) Capital goods -- Construction.}
\label{array500} 
\end{figure}
In Fig.~\ref{array500} we show a color representation of the categorical variables of some of the stocks included in the Bonferroni network. Specifically, we consider stocks belonging to the 10 largest clusters detected by the Infomap algorithm. In Fig.~\ref{array500} a green spot indicates a state \emph{up} of the excess return whereas a red spot indicates a state \emph{down}. The state \emph{null} of the excess return is represented by a black spot. Vertical blue lines are separating different clusters. Horizontal lines of the same color indicates co-occurrence of a state at a specific trading day. The first cluster is the largest cluster of Services-Real estate operations, whereas the second cluster  is the Technology cluster. From Fig.~2 of the main text we know that these two clusters present opposite movements of the excess return, and in fact Fig.~2 of this Supplementary Information shows that for these clusters categorical variables take opposite states in many days. Figure \ref{array500} also shows that days of state co-occurrence are continuously present over the entire investigated time period and therefore the interrelations among different groups of stocks are continuously observed for long periods of time. 

\section{Network of movies}\label{cdsvn}

The largest component of the adjacency movie network comprises 77,193 movies whereas the second largest component has only 11 movies.
When we apply the Infomap partitioning algorithm to the unweighted adjacency movie network we obtain a partitioning of the network which presents 2,451 distinct clusters. The cluster size decreases smoothly from the largest value of 13,608 down to the smallest value of 2. We will see in the following discussion that the partitioning of the network presents a certain degree of informativeness about the system. In fact the obtained clusters present a certain degree of homogeneity with respect to the main country of production, the language and some classes of genre of movies.

The FDR network is characterized by a largest connected component of 30,934 movies. The Infomap algorithm makes a partition of this and other components of the network into 3,967 clusters whose size is decreasing from 1,478 to 2 movies. Table 1 of the main text shows that the Bonferroni network does not present a giant connected component. In fact the largest connected component comprises only 13\% of the movies linked in the network. However, the application of the Infomap algorithm refines the natural partitioning of the network by detecting 2,782 clusters whose size is ranging from 577 to 2 movies. We also note that the number of connected components in the Bonferroni network (2,456) is roughly equal to the number of the Infomap clusters in the adjacency network (2,451).

\subsection{Community detection in weighted movie networks}\label{wn}

Our method provides a full control of the statistical validation of links against a random null hypothesis taking into account the fact that different movies have a different number of actors. The system also presents a second source of heterogeneity. In fact, different actors typically play a different number of movies. In our sample the number of movies played by a single actor is ranging from 1 to 247.
As discussed in the main text, we do not have a rigorous and computationally feasible way to also take into account this second source of heterogeneity in our statistical validation procedure. We therefore use a heuristic approach, and take into account this heterogeneity by following Reference\footnote{Newman, M. E. J. {\it Scientific collaboration networks. II. Shortest paths, weighted networks, and centrality.} {\em Phys. Rev. E} {\bf 64} 016132 (2001)}.  
We adapt the procedure proposed in Reference$^6$ to our system by weighting the link present between movie $a$ and movie $b$ with a value $w_{ab}$ that takes into account the number of movies played by the actors playing both the movies. Specifically, %
\begin{equation}
w_{ab}=\sum_{i=1}^Q \frac{1}{N_i-1},
\label{weight} 
\end{equation}
where the sum is taken over all the $Q$ actors who play both movie $a$ and $b$, and $N_i$ is the total number of movies played by actor $i$.

\begin{figure}
\includegraphics[scale=0.3,angle=0]{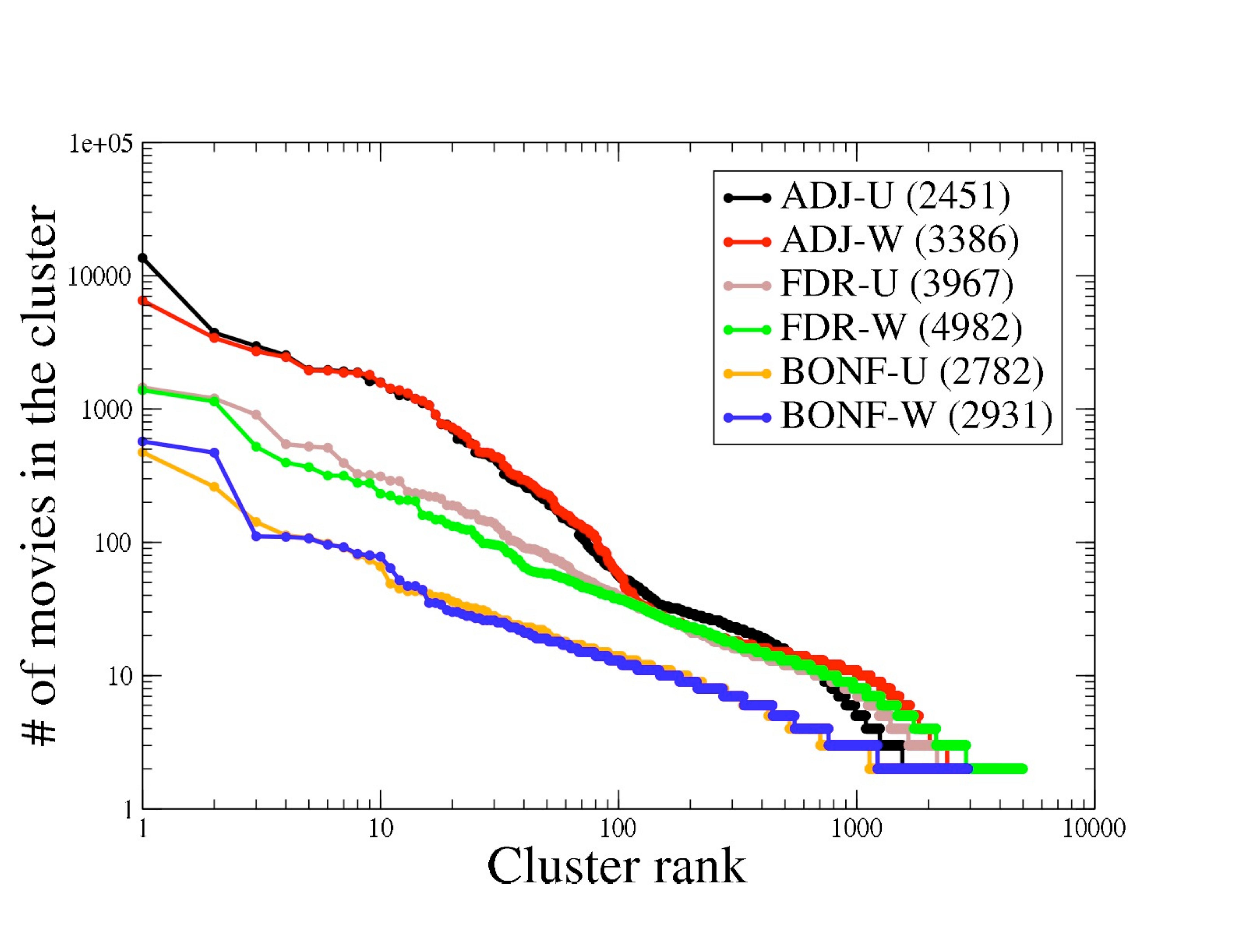}
\caption{Rank plot of the size of clusters obtained with the Infomap algorithm for the adjacency movie network, the FDR network and the Bonferroni network both for the unweighted and weighted links. The difference between the partitions decreases for the statistically validated networks (see text for a measure of the mutual information between unweighted and weighted partitions). In the legend, the number in parenthesis is the number of detected clusters in the corresponding network.}
\label{ClusterSize} 
\end{figure}

By performing community detection on the weighted adjacency movie network, we obtain a more refined partitioning of the 78,686 movies present in the network. Specifically, the clusters obtained with the Infomap algorithm present 3,386 clusters whose size is decreasing from 6,523 to 2. By performing community detection on the weighted statistically validated networks, we obtain results that are very similar to those obtained for the corresponding unweighted networks.
The impact of considering link weights in community detection is quantitatively discussed in the following subsection.

\subsection{How link weights affect the community structure of networks}
The Infomap method allows to take into account link weights. This feature implies that results of community detection in a given network may significantly change when weights of links are considered. For each network, we quantify the difference between the partition obtained without using link weights and the partition obtained by taking link weights into account by calculating the (normalized) mutual information between the two partitions\footnote{Danon, L., Diaz-Guilera A., Duch J., \& Arenas A. Comparing community structure identification. {\em J. Stat. Mech.-Theory Exp.} P09008 (2005)}. The mutual information takes the maximum value of 1 for two identical partitions of the network. We observe a value of 0.798, 0.913 and 0.976 for the adjacency, FDR and Bonferroni networks, respectively. We therefore observe a net increase of the mutual information when we consider statistically validated networks. Furthermore, the mutual information reaches a value very close to 1 for the Bonferroni network, which is obtained under the most restrictive statistical requirements. In other words, we observe that the source of heterogeneity of the actors' productivity has a minor impact into the partitioning of the statistically validated networks, especially for the Bonferroni network.

\subsection{Cluster size and inclusiveness}
In the following, we separately discuss the results obtained for the partitioning of the adjacency, FDR and Bonferroni weighted networks. Results obtained for the Bonferroni network are rather similar to those obtained for the FDR network.  The size profile of the clusters obtained by partitioning the adjacency, FDR and Bonferroni networks are shown in Fig.~\ref{ClusterSize}, both in the case of unweighted and weighted networks. It is quite clear from the figure that the cluster size decreases from the largest to the smallest cluster in a pretty different way for the adjacency networks and the statistically validated networks. In fact, the FDR and Bonferroni networks present a decay of cluster size versus its rank that is well approximated by a power-law decay. 

In the majority of cases, the clusters detected in the Bonferroni network correspond to the strongest interconnected parts of larger clusters detected in the FDR network. The clusters of the FDR network correspond in turn to sets of movies which in large majority are present in bigger clusters observed in the weighted adjacency network. This general observation should not be seen as a strict inclusive relation but we wish to point out that a sort of ``typical" inclusiveness is observed for most of the detected clusters.   

In the next subsection we describe results about the characterization of clusters in the different networks. 

\subsection{Cluster characterization}\label{cluster}

We analyze the clusters of movies we obtain for the three different weighted networks, by considering the over-expression of specific characteristics of the movies contained in each cluster. By using the information about movies as provided by IMDb, we consider 4 different classifications of movies. Indeed the IMDb reports for each movie indication about (i) country or countries of production, (ii) language or languages used in the movie, (iii) movie genre (or genres) and (iv) location or locations where the movie was shot. Only in a limited number of cases some of these informations are not available. When this happens we indicate the missing attribute about the movie as ``not available". We characterize clusters obtained for all the networks by separately testing the over-expression of each attribute present in each one of the above mentioned 4 classifications.   

In different networks we observe a different profile of over-expression. The degree of specificity is higher for smaller clusters and therefore a higher specificity is observed for the Bonferroni and for the FDR networks. This is especially true for the genre and the location classifications. In general, the country and language over-expression is quite specific for most clusters in the investigated networks. Exceptions are clusters containing movies produced in former Yugoslavia and Soviet Union. This is due to the fact that during the investigated period these countries have split into several independent countries.

In Table~\ref{summary_oe} we summarize the number of over-expressions observed in the clusters identified by the Infomap in the weighted networks. We also report in parenthesis the number of distinct clusters where at least one over-expression has been observed. By comparing the number of over-expressions with the number of characterized clusters, one can estimate the average number of over-expressions per cluster. This number decreases, for any considered classification, when we move from the weighted adjacency network to the weighted FDR network and then to the weighted Bonferroni network (the only exception being observed for the language characterization when moving from the FDR to the Bonferroni network). For example, in the case of the genre over-expression, the average number of over-expressions per cluster is 1.41, 1.34 and 1.33 for the adjacency, FDR and Bonferroni network, respectively. The decrease is more pronounced for the genre and filming location characterization. This observation quantitatively indicates a higher specificity in cluster characterization for the statistically validated networks.
\begin{table}
\caption{Summary of the over-expression of production country, language, genre and filming location observed in the clusters obtained by performing the Infomap partitioning of the adjacency weighted movie network (ADJ-W), FDR weighted movie network (FDR-W) and the Bonferroni weighted movie network (BONF-W). For each of the four considered classifications, we report the total number of observed over-expressions for each network. The number in parenthesis is the number of distinct clusters where at least one over-expression has been observed.}
\begin{tabular}{|l|c|c|c|}
\hline
~ & ADJ-W &  FDR-W &  BONF-W \\
\hline
movies in all clusters & 78,686 & 37,429 & 12,850\\
number of clusters & 3,386 & 4,982 & 2,931 \\
\hline
Production country & 1,206       &  1,944 & 1,009\\  
over-expression      & (1,115)    & (1,816) & (960)\\
\hline
Language                 & 601     & 1,429 & 819\\
over-expression      & (494)   & (1,297) & (729) \\
\hline
Genre                        & 629       & 715 & 373 \\
over-expression      & (445)    & (533) & (281)\\
\hline
Filming location      & 2,196     & 1,836 & 853 \\
over-expression     &  (793)    & (1,123) & (571)\\
\hline
\end{tabular}
\label{summary_oe}
\end{table}
In the next section, we comment in detail two specific cases, in order to illustrate some of the changes of sensitivity and specificity in the over-expression characterization of clusters in the different weighted networks. 

\subsection{Case studies}\label{comp}

\subsubsection{Largest cluster of the weighted adjacency network and the overlapping FDR clusters}

We discuss the case of the largest cluster observed in the partition of the adjacency weighted network. This is a cluster of 6,523 movies mainly in English and mainly produced in the USA. In the caption of Table~\ref{adjw1}, we report the major over-expressions observed for this cluster. The over-expressed production country is USA and in fact 6,433 movies of the cluster have been produced or co-produced in that country. The over-expressed languages are English and Vietnamese. However, it should be noted that the number of movies filmed in these languages is quite different. In fact there are 6,264 movies where the language is English and only 17 movies where the language is  Vietnamese. The over-expression of Vietnamese is observed because only 64 movies in Vietnamese are present in the weighted adjacency network. The genre over-expression involves 14 different genres and the filming location over-expression involves 97 different locations. The large majority of filming locations are in California but cities of many other states are also observed.

We observe that 3,600 movies of the above described adjacency network cluster
are split in many clusters of the weighted FDR network. In Table~\ref{adjw1} we report some information about the seven FDR nework clusters having the largest intersection with the considered adjacency network cluster. Cluster labels in the figure are those provided by the Infomap. The complete list of clusters and movies for all the networks is available upon request to the authors. From the Table it is evident that the size of the FDR clusters is more than one order of magnitude smaller than the size of the  adjacency cluster. In other words the Infomap partitioning of the FDR network is quite refined for USA movies. These clusters of movies are almost always characterized by USA as production country and English as  language. The genre and filming location over-expression provide the main characterization of FDR clusters. Table~\ref{adjw1} shows that the FDR cluster 17 is mainly a cluster of animation movies and, of course, for this cluster no filming location over-expression is observed except the indication of  not available (NA). Cluster 56 is a cluster of action movies and the main filming location of them is Los Angeles, CA. Clusters 47, 91 and 123 are all clusters of comedy movies and the over-expressed filming location is again Los Angeles, CA. Cluster 188 is composed by a group of horror movies and a group of science fiction movies, while cluster 95 mainly includes thriller and crime movies. Interestingly, for this last cluster the over-expressed filming locations are all in Florida. This case study shows that the FDR network loses in sensitivity with respect to the adjacency network (less movies are involved in the FDR network) but significantly gains in specificity (clusters in the FDR network are more homogeneous, especially with respect to the genre and filming location characterization). To provide a further example of this improvement in specificity it is worth noting that 8 of the 17 movies in Vietnamese language present in the largest cluster of the weighted adjacency network are found in cluster 775 of the FDR network, which is composed by only 15 movies. 

\begin{sidewaystable}
\begin{small}
\caption{Over-expression of production country (C), language (L), genre (G) and filming locations (F) for seven largest clusters of the FDR weighted network. Here we consider only  those movies that are also present in cluster 1 of the adjacency weighted network (ADJ-W). In fact, the number in parenthesis indicates the number of movies in a specific FDR-W cluster that are also present in cluster 1 of the adjacency weighted movie network. The cluster 1 of the adjacency weighted network is composed by 6523 movies. The over-expressions that characterize such cluster are:   Production country - USA (6344 movies); Language - English (6264) and Vietnamese (17); Genre -  Comedy (2385), Thriller (1300), Action (1001), Romance (707), Crime (666), Horror (491), Family (441) Adventure (425), Sci-Fi (394), Fantasy (313), Animation (293), Mystery (284), Sport (125) and Western (70); Filming location - Los Angeles, CA (2459), San Francisco, CA (159), Pasadena, CA (141), Santa Clarita, CA (136), Las vegas, NV (135), Long Beach, CA (129), Culver City, CA (121), Burbank, CA (115), California (103), Santa Monica, CA (84) and other 87 locations. }
\begin{tabular}{|c|c|c|c|c|c|c|c|}
\hline
cluster ID               & FDR-W 17& FDR-W 56 & FDR-W 188 & FDR-W  47 & FDR-W  91 & FDR-W  123 & FDR-W 95 \\
movies in cluster  & 98 (87)      & 59 (58)       & 48 (46)         & 46 (41)         & 53 (39)       & 45 (37)           & 45 (33) \\
 \hline
C & USA 88           & USA 58                & USA 47     & USA 44               & USA 53               & USA 45              & USA 44 \\
 \hline
L & English 93       &  English 57          &   English 46 &  English 46            & English 49           & English 43         & English 45\\
 \hline
G & Animation 77  &  Action 47            &  Horror 12   &  Comedy 38          &  Comedy 33        &  Comedy 34       &  Thriller 18\\
G & Family 62        &  ~                            &  Sci-Fi 12    &  ~                              &  Music 6               & ~                          &  Crime 14 \\
G & Adventure 45 &  ~                            & ~                  &  ~                            &  ~                          & ~                          &  \\
G & Fantasy 29     &  ~                             & ~                  & ~                             & ~                          & ~                            &  \\
G & Sci-Fi 15         &  ~                             & ~                  & ~                             & ~                            & ~                          & \\
G & Musical 9        &  ~                            & ~                   & ~                            & ~                             & ~                         &  \\
 \hline
F & NA 75              & Los Angeles 19  & ~                   & Los Angeles 30    & Los Angeles 21    & Los Angeles 25  & Miami 23 \\
F & ~                       & ~                            & ~                  & Long Beach 5      & ~                             & ~                           & Miami Beach 9\\
F & ~                       & ~                             & ~                  & ~                              & ~                            & ~                          & Fort Lauderdale FL 6\\
F & ~                       & ~                             & ~                  & ~                              & ~                             & ~                          & Florida 5\\
F & ~                       & ~                             & ~                  & ~                             & ~                             & ~                           & Coral Gables FL 4\\
F & ~                       & ~                             & ~                  & ~                              & ~                             & ~                          & Hollywood FL 3\\
F & ~                       & ~                            & ~                  & ~                              & ~                             & ~                           & Key Biscayne FL 3\\
\hline
\end{tabular}
\label{adjw1}
\end{small}
\end{sidewaystable}

\subsubsection{A cluster of Indian movies of the weighted adjacency network and the overlapping clusters of the FDR and Bonferroni networks}

The second case study concerns a cluster of Indian movies. The weighted adjacency movie network presents five large clusters of Indian movies. Here we discuss the properties of the second largest cluster of these five Indian movies clusters. We do not consider the largest one, because it is already very homogeneous according to the language: 90\% of indian movies in this cluster are in Hindi. We indicate the selected cluster of Indian movies as cluster 24 of the weighted adjacency network. This cluster consists of 648 movies. In the caption of Table~\ref{adjw24} we report all the over-expressions observed for this cluster. 
The over-expressed production country is indeed India and over-expressions are also observed for four languages spoken in India, five distinct movie genres, and eight filming locations. By comparing the clusters of the weighted FDR network with this cluster of the weighted adjacency network, we observe a large overlapping of movies. For example, 515 movies of cluster 24 are present in clusters 5 and 43 of the weighted FDR network. In other words, the Indian cluster 24 detected in the weighted adjacency network splits into two distinct clusters in the weighted FDR network.
The first cluster (cluster 5) comprises movies where the language spoken is mainly Telugu, whereas the second cluster (cluster 43) mainly comprises movies in Tamil. The characterization of genre and filming location is more specific than the one observed for cluster 24 of the adjacency network, but the degree of specificity is not too high (see the first two columns of Table~\ref{adjw24}). 

A higher degree of specificity is observed when we consider the clusters of the weighted Bonferroni network. In Table~\ref{adjw24}, we show the over-expression characterization of the five largest clusters of Bonferroni network overlapping with cluster 24 of the weighted adjacency network (last five columns of Table~\ref{adjw24}). There is a unique language characterization per cluster at this level. The filming location characterization is poor due to the fact that this information is often absent for Indian movies recorded in the IMDb. In fact the ``not available" (NA) over-expression is the most frequent one. 

In conclusion, we also notice for Indian movies the ability of statistically validated networks to describe communities of movies that are smaller but more homogeneous, according to the considered classifications, with respect to the communities of movies in the adjacency network.

\begin{sidewaystable}
\caption{Over-expression of production country (C), language (L), genre (G) and filming locations (F) for two largest clusters of FDR weighted network and five largest clusters of Bonferroni weighted networks. Here we consider the movies that are also present in cluster 24 of the adjacency weighted movie network. In fact, the number in parenthesis indicate the number of movies in a specific FDR-W or BONF-W cluster that are also present in cluster 24 of the adjacency weighted movie network. The cluster 24 of the adjacency weighted network is composed by 648 movies. The over-expressions that characterize such cluster are:  Production country - India (643 movies); Language - Telugu (438), Tamil (196), Hindi (52) and Kannada (14); Genre -  Drama (312), Action (213), Romance (180), Family (53) and Musical (48); Filming location - Not Available (515), Hyderabad (53), Chennai (31), India (17), Andhra Pradesh (8), Rajahmundry (5), Tamil Nadu (4) and Vikarabad (3).}
\begin{tabular}{|c|c|c|c|c|c|c|c|}
\hline
cluster ID & FDR-W 5      & FDR-W 43  & BONF-W 10  & BONF-W 13 & BONF-W 309  & BONF-W 607 & BONF-W  806 \\
~               & 396 (390)          & 132 (125)      & 111 (111)        & 110 (110)        & 13 (13)          & 10 (10)          & 15 (13)   \\
 \hline
C & India 395                   & India 132         & India 111     & India 110                 & India 13      & India 10  & India 15 \\
\hline
L & Telugu 375                & Tamil 120         & Telugu 111  & Telugu  109            & Telugu 12 & Telugu 9 & Tamil 14 \\
L & Tamil 15                     & Hindi 21            &  ~                   &  ~                              & ~                  & ~                 & ~ \\
L & ~                                  & Telugu 21         &   ~                  & ~                                & ~                & ~               & ~ \\
\hline
G &  Action 132                & Romance 52  & Family 25      & Action 47                 & ~                & ~                 & ~ \\
G & Romance 94             &Action 48         &  Musical 11   &  Romance 39           & ~                 & ~                 & ~ \\
G & Family 40                  & Musical 17      & ~                     & ~                                & ~                & ~                 & ~ \\
G & Musical 24                &  ~                      & ~                     & ~                                 & ~                & ~                & ~ \\
\hline
F & NA 315                      & NA 99                 & NA 98           & NA 78                        & ~                 & ~                & ~ \\
F & Hyderabad 45          & Chennai 15      & ~                    & Hyderabad 13         & ~                & ~                 & ~ \\
F &  Andhra Pradesh 7  & Tamil Nadu 3   & ~                    & Andhra Pradesh 5   & ~                 & ~                 & ~ \\
F &  Rajahmundry 4       & ~                         & ~                    & ~                                 & ~                & ~                 & ~ \\
\hline
\end{tabular}
\label{adjw24}
\end{sidewaystable}

\end{document}